\documentclass[a4paper,12]{spieman}  
\usepackage[]{graphicx}
\usepackage[]{amsmath}
\usepackage{units}
\usepackage{paralist}
\usepackage{textcomp}
\usepackage{subfig}
\usepackage{tabularx}
\usepackage{booktabs}
\usepackage{rotating}
\usepackage{multirow}
\usepackage{xspace}
\usepackage{float}
\usepackage[symbol]{footmisc}
\usepackage{lineno}

\DeclareMathAlphabet\mathbfcal{OMS}{cmsy}{b}{n}

\title{Hyperspectral Camera Selection for Interventional Health-care} 

\author[a]{Anant S. Vemuri}
\author[a]{Sebastian Wirkert}
\author[a]{Lena Maier-Hein}
\affil[a]{Division of Computer Assisted Medical Interventions, DKFZ, Heidelberg}

\begin{document} 
\maketitle 
\begin{abstract}

Hyperspectral imaging (HSI) is an emerging modality in health-care applications for disease diagnosis, tissue assessment and image-guided surgery. Tissue reflectances captured by a HSI camera encode physiological properties including oxygenation and blood volume fraction. Optimal camera properties such as filter responses depend crucially on the application, and choosing a suitable HSI camera for a research project and/or a clinical problem is not straightforward. We propose a generic framework for quantitative and application-specific performance assessment of HSI cameras and optical subsystem without the need for any physical setup. Based on user input about the camera characteristics and properties of the target domain, our framework quantifies the performance of the given camera configuration using large amounts of simulated data and a user-defined metric. The application of the framework to commercial camera selection and band selection in the context of oxygenation monitoring in interventional health-care demonstrates its integration into the design work-flow of an engineer. The advantage of being able to test the desired configuration without the need for purchasing expensive components may save system engineers valuable resources.



\end{abstract}



{\noindent \footnotesize\textbf{*}Anant S. Vemuri (\linkable{a.vemuri@dkfz-heidelberg.de}), Lena Maier-Hein (\linkable{l.maier-hein@dkfz-heidelberg.de})}

\begin{spacing}{1.2}   
\section{INTRODUCTION}
\label{sec:intro}  
Hyperspectral imaging (HSI), also sometimes referred to as imaging spectroscopy \cite{wolfe1997introduction}, finds its roots in remote sensing. Medical HSI on the other hand is an emerging imaging modality for health-care applications, especially in disease diagnosis and image-guided surgery. HSI acquires a three-dimensional dataset called a hypercube, that contains two spatial dimensions and one spectral dimension, which provides reflectance measurements for each point in the image at multiple wavelengths of light. During the progression of a disease \cite{costas2011multi}, the absorption, fluorescence and scattering characteristics of tissue change, which are encoded in the reflectances. Hence, these measurements from a HSI camera carry quantitative diagnostic information about tissue pathology. The knowledge about tissue parameters such as oxygenation (SaO$_2)$ and blood volume fraction (BVF) has motivated a considerable body of research on early cancer detection \cite{claridge2013model}, image-guided therapy involving bowel anastomosis \cite{wirkert2016robust}, hemodynamic response assessment of human cortex \cite{pichette2016intraoperative} and diagnosis of cervical neoplasia \cite{ferris2001multimodal}. The various applications presented in literature cover the ultraviolet (UV), visible (VIS), and near-infrared (near-IR or NIR) regions \cite{lu2014medical}.

No approach has been presented in literature thus far, that is able to analyze and quantify the performance of an HSI camera setup for clinical scenarios. To address the gap from theory to application, this paper presents a generalized framework for quantitative and application-specific performance assessment of HSI cameras without the need for a physical system setup. The following components can alter the performance of an optical system setup: \begin{inparaenum}[a)]
	\item The quantum efficiency of the image sensor,
	\item filter responses for the various bands employed,
	\item light source characteristics,
	\item effects of additional application-specific components such as optic fiber, glass lens (for example in surgery), and 
	\item signal-to-noise ratio (SNR) or contrast-to-noise ratio (CNR) in the target application depending on for eg. the chosen imaging hardware, limitations in lighting conditions.
\end{inparaenum}

The proposed framework leverages simulation of light transport in tissue, using existing software such as the Monte Carlo modeling of light transport in multi-layered tissues \cite{wang1992monte} (MCML) to generate spectral reflectances which can be combined with the above mentioned parameters to obtain realistic camera reflectances. Given a user-defined performance metric, the simulated data can then be used to quantify the performance of an HSI setup offline or to fine-tune a given optical system for best efficiency in a target application without the need for physical hardware.

Sec. \ref{sec:method} describes our framework in detail. Sec. \ref{sec:experiments_and_results} presents the experiments to verify and validate our software framework and two sample use-cases that demonstrate how the proposed framework can be used in practice. Finally, Sec. \ref{sec:discussion} offers a perspective on usage of the framework to other use-cases in hyperspectral applications in health-care.

\section{Method}
\label{sec:method}





\begin{figure}
\centering
  \includegraphics[width=0.98 \columnwidth]{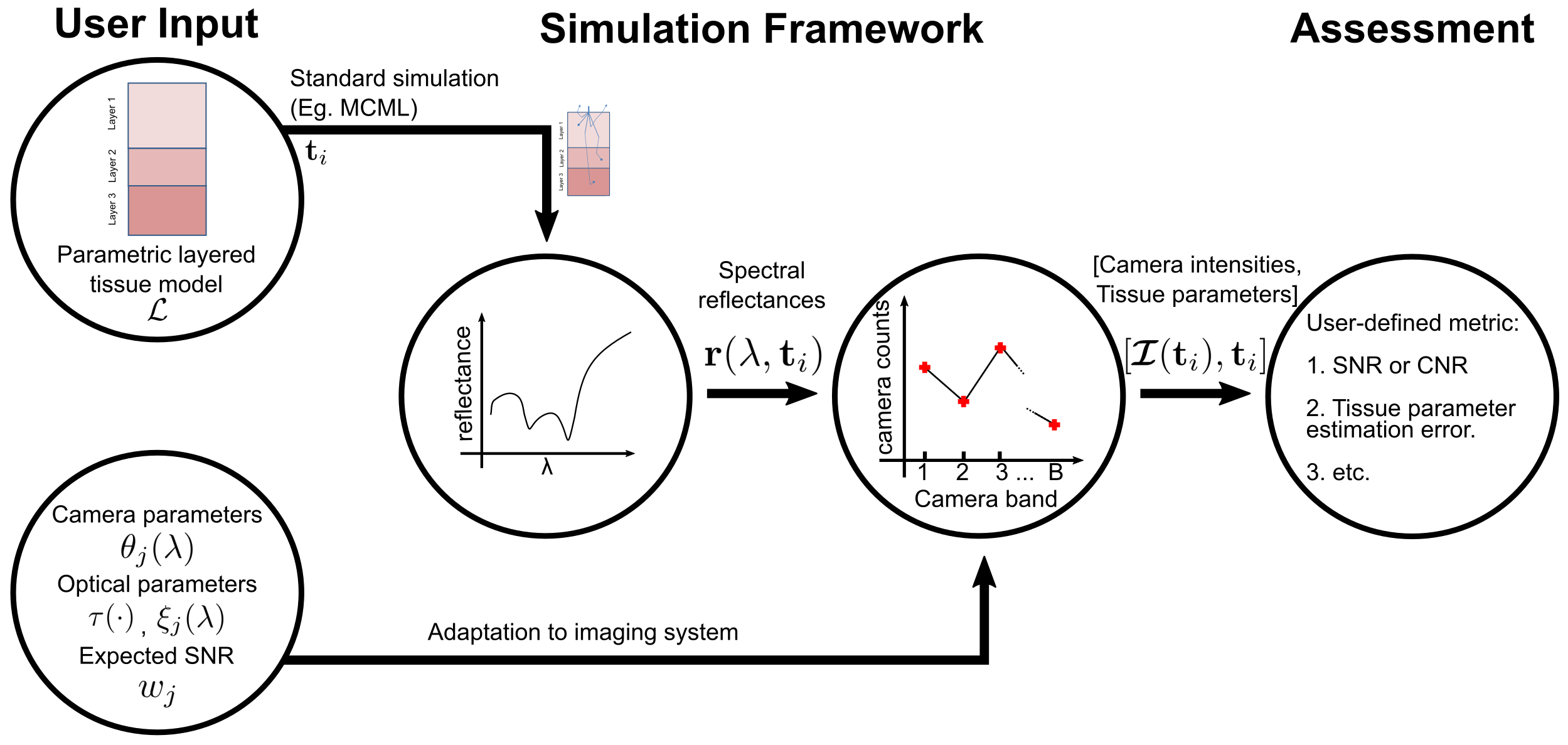}
  \caption{Overview of our camera assessment framework.  A tissue sample $\textbf{t}_i$ is modeled as a layered structure with each layer characterized by a set of seven optically relevant tissue properties, as summarized in Tab. \ref{tab:parameter_description}. Based on Monte-Carlo methods, realistic tissue reflectances $\mathbf{r}(\lambda, \mathbf{t}_i)$ can be generated . The reflectances are transformed to (virtual) camera space using information on the camera parameters $\theta_j$, optical parameters $\xi_j(\lambda)$, camera response ($\tau(\cdot)$) and signal-to-noise ratio ($w_j$) (each indexed by $j^{\text{th}}$ camera band). The output of the simulations is the pair $[\mathbfcal{I}(\mathbf{t}_i), \mathbf{t}_i]~\forall i\in[1,N]$, which the user can use with metric to quantify system performance.} \label{fig:flowchart_generative_model}
\end{figure}

This paper builds upon the following scientific contributions: \begin{inparaenum}[(a)]
    \item {Parametric models to represent human tissue (cf. e.g. \cite{jacques2013optical, boas2016handbook,hidovic2005modelling}),}
    \item {publicly available software for simulating light transport through tissue (e.g. MCML \cite{alerstam2010next,wang1992monte} and Geant4 \cite{allison2016recent}) and,}
    \item {mathematical model for transforming simulated reflectances to camera space.}
\end{inparaenum} 
Inspired by this previous work, we propose an end-to-end evaluation framework that assesses the performance of a camera in an application-specific manner. The assessment relies on the following prior information to be provided by the HSI engineer (cf. Fig. \ref{fig:flowchart_generative_model}): 

\begin{enumerate}
	\item  \textit{User input on the camera and the optical subsystem} is represented by parameters related to the filter, transmission characteristics of the optics (lens, filters and optic fibers), light source characteristics, the expected SNR from the target domain etc.
	
	\item \textit{User input on the target domain} is formally described by instantiating a layered tissue model (\ref{sec:layered_tissue_model}) that has been introduced in previous work \cite{wirkert2016robust}. The user may choose not to instantiate a tissue-specific model or use a generic tissue model\cite{wirkert2017physiological} instead.
	
\end{enumerate}


	
Based on this input, the \textit{generative model} is used to build a large set of simulated camera reflectances from the target domain, as described in  Sec.~\ref{sec:generative_model}. A user-defined \textit{metric} is then used to quantify the performance of the optical system setup based on the simulated data as shown in Sec. \ref{sec:metric_computation}. 


\subsection{Generative model}
\label{sec:generative_model}
The forward model consists of three parts: \begin{inparaenum}[(1)]
\item A (generalized or specific) layered tissue model that formally describes the composition of possible tissues (Sec. \ref{sec:layered_tissue_model}).
\item Realistic simulation of spectral reflectances, computed using the tissue model (Sec. \ref{sec:simulated_spectral_reflectances}).
\item Transformation of the simulated reflectances to the camera space using the camera characteristics (Sec. \ref{sec:adapting_imaging_system}), provided by the user.
\end{inparaenum}

\subsubsection{Layered tissue model}
\label{sec:layered_tissue_model}

The tissue is modeled as a layered structure \cite{wirkert2016robust} $(\mathcal{L} = [\mathbf{l}_1, ... \mathbf{l}_K])$, with each layer $\mathbf{l}_k \in \mathcal{L}$, characterized by a set of optically relevant tissue properties: 
$\mathbf{l}_k = \{ v_{\text{hb}, k}, s_k, a_{\text{mie}, k}, b_k, g_k, n_k, d_k\}$, detailed in Tab. \ref{tab:parameter_description}. Three assumptions are made in this paper about the layered structure; First, hemoglobin is the major chromophore in the tissue. Second, as suggested by Claridge et al.\cite{claridge2013model}, $s_k$ is assumed to be constant across layers, which is a reasonable assumption since, the layers share a common blood supply (and will henceforth be referred to as $s$). Third, pixel independence is implicitly assumed, by modeling tissue as homogeneous, infinite slabs. This leads to fewer assumptions on the 3D composition of tissue (e.g. vessel structures), but prevents modeling cross-talk between pixels.



\begin{figure}[h!]
	\centering
	\includegraphics[width=0.8 \columnwidth]{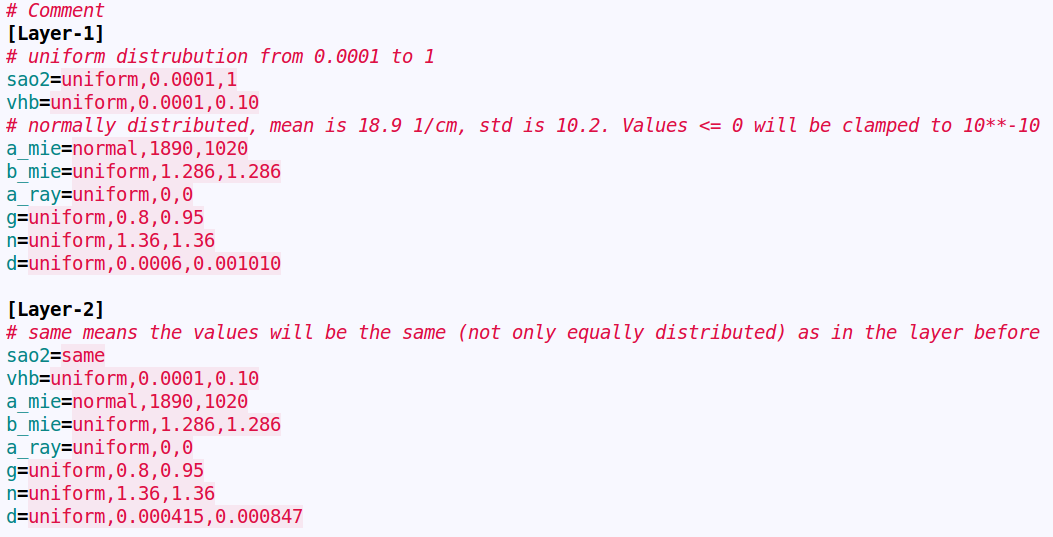}
	\caption{Typical configuration file structure for a two-layer tissue model.}\label{fig:tissue_ini_sample}
\end{figure}

The parameters of the layered model described in Tab. \ref{tab:parameter_description} are included into a configuration file, which is input to the proposed assessment framework. Fig. \ref{fig:tissue_ini_sample} presents a sample configuration file for a three layered tissue. Each layer parameter has three values; the distribution type (uniform, normal, step etc.), followed by the upper and lower bound for the uniform distribution (mean and std. dev for normal distribution). A special case, when the layer parameter replicates the previous layer, the keyword ``same'' is used. This simply copies the value from previous layer's corresponding parameter. From the layered tissue model presented in Tab. \ref{tab:da_generic}; each parameter is drawn randomly from the defined ranges to form a tissue instance $\mathbf{t}_i \in \mathbf{T}$ $(\forall i \in [1,N])$. Each tissue instance is thus represented by $K \cdot 6 + 1$ parameters (where the $+1$ represents the oxygenation $s$, which is assumed constant across all layers).

\begin{table}[!]
	\centering
	\caption{Parameters used to instantiate a single layer of tissue model as described Sec. \ref{sec:layered_tissue_model}. \label{tab:parameter_description}}
	\begin{tabular}{lp{12.7cm}l}  
		\toprule
		Parameter            & Description  & Units\\
		\midrule
		$v_{\text{hb}}$      & Blood volume fraction (BVF), defined as the amount of blood occupying a unit volume of tissue &  $[\%]$    \\
        \noalign{\smallskip}
		$s$                  & Tissue oxygenation, defined as the amount of oxygen-saturated hemoglobin relative to total hemoglobin (unsaturated + saturated) in the blood &  $[\%]$\\
        \noalign{\smallskip}
		$a_{\text{mie}}$     & Reduced scattering coefficient at 500$\unit{nm}$ $[\unit{cm}^{-1}]$ to represent the amount of scattering in the tissue as proposed in \cite{jacques2013optical} & $[\unit{cm}^{-1}]$\\
        \noalign{\smallskip}
		$b$                  & Scattering power, which characterizes the exponential wavelength dependence of the scattering \cite{jacques2013optical} & [-]\\
        \noalign{\smallskip}
		$g$                  & Anisotropy factor, characterizing the direction of scattering \cite{jacques2013optical} & [-] \\
        \noalign{\smallskip}
		$n$                  & Refractive index of the tissue layer & [-]\\
        \noalign{\smallskip}
		$d$                  & Layer thickness. &  $[\unit{\mu m}]$\\
		\bottomrule
	\end{tabular}
\end{table}

\begin{table}[!]
        \centering
        
        \caption{The simulated ranges of physiological parameters, and their usage in the simulation set-up as described above.\label{tab:da_generic}}
        
        \begin{tabularx}{\textwidth}{@{}XXXXXXXX@{}}
            \toprule
            & $v_{\textrm{hb}} \lbrack \% \rbrack$           & $s \lbrack \% \rbrack$  & $a_{\text{mie}} \lbrack \unit{\frac{1}{cm}} \rbrack$ & $b_\text{mie}$         & $g$      & $n$         & $d \lbrack\unit{mm}\rbrack$\\ \midrule
            layer 1-3: & 0-30 & 0-100 & 5-50 & .3-3 & .8-.95 & 1.33-1.54 & .02-2\\ \midrule
            \multicolumn{8}{l}{$\mu_a(v_{\text{hb}}, s, \lambda) = v_{\text{hb}} (s \epsilon_{\text{HbO2}}(\lambda) + (1-s)\epsilon_{\text{Hb}}(\lambda)) \ln (10)(150\unit{gL}^{-1})(64,500\unit{g}\,\unit{mol}^{-1})^{-1}$}\\
            \multicolumn{8}{l}{$\mu_s(a_{\text{mie}}, b, \lambda) = \frac{a_{\text{mie}}}{1-g}(\frac{\lambda}{500\unit{nm}})^{-b}$}\\ \midrule
            \multicolumn{8}{l}{simulated samples: 500K}\\
        \end{tabularx}
\end{table}

\begin{table}[!]
    \centering
    \caption{Factors influencing the camera intensity. measurements.\label{tab:camera_measurement_factors}}
    \begin{tabular}{lp{14cm}}  
        \toprule
        Parameter                     & Description \\
        \midrule
        
        $\mathbf{r}(\lambda, \mathbf{t}_i)$             & Spectral reflectance for the tissue $\mathbf{t}_i$ at wavelength $\lambda$ discretized by $\lambda\in\Lambda$  \\
        \noalign{\smallskip}
        \multirow{2}{*} {$\xi_j(\lambda)$}              & Represents the irradiance of the illuminant (e.g., Xenon or Halogen) and other aspects in the imaging system such as, transmittance \cite{styles2006quantitative} of optical system. Some of these factors may depend on the position in image $\mathbf{p}$, but will be ignored in this paper. The vector form is represented by $\Xi$\\ 
        \noalign{\smallskip}
        \multirow{2}{*} {$\theta_j(\lambda)$}           & This variable encompasses the other linear factors affecting the formation of the image. This includes (linear) parameters of the HSI camera affecting the response of band $j \in [1,\mathbf{B}]$ and the quantum efficiency of the camera. The vector form is represented by $\Lambda$\\ 
        \noalign{\smallskip}
        \multirow{2}{*} {$\tau_j(\cdot)$}               & The function relating scene radiance to image brightness is called the camera response, often intentionally non-linear to cover a higher range of irradiances within the possible image intensities. \cite{grossberg2003space}. However for scientific cameras it is linear in the operating range\\
        \noalign{\smallskip}
        \multirow{2}{*} {$w_{j}$}                       & Camera noise for band $j$, mainly coming from (1) shot noise due to the particle nature of light and (2) dark noise, which is dependent on sensor temperature and linear in integration time \cite{healey1994radiometric}\\ 
        \bottomrule
    \end{tabular}
\end{table}

\subsubsection{Simulated spectral reflectances}
\label{sec:simulated_spectral_reflectances}



The tissue configuration presented in the previous section, is used to compute spectral reflectances\cite{wirkert2016robust}. For a tissue $(\mathbf{t}_i)$ drawn from a layered model $(\mathcal{L})$, reflectance spectrum $\mathbf{r}(\lambda, \mathbf{t}_i)~\forall \lambda \in \Lambda$\footnote[3]{sample wavelength range $(\Lambda)$: 300-1000nm (stepsize=2nm)} is computed by evaluating a function $\mathbf{r}_{\text{sim}}$ at wavelengths $\lambda$ as,



\begin{equation}
\mathbf{r}(\lambda, \mathbf{t}_i) = \mathbf{r}_{\text{sim}}(\mathbf{\lambda}, \mathbf{t}_i)
\end{equation}



MCML\cite{wang1992monte} is used to evaluate $\mathbf{r}_{\text{sim}}(\mathbf{\lambda}, \mathbf{t}_i)$ in this paper, because Monte-Carlo (MC) models are widely considered to be the gold standard for calculating how light travels through biological tissue \cite{zhu2013review}. In MCML tissue is modeled as a set of infinite, homogeneous slabs. Each slab is characterized by its depth $(d)$, anisotropy $(g)$, absorption coefficient $(\mu_a)$, scattering coefficient $(\mu_s)$ and refractive index $(n)$ (as presented in Tab. \ref{tab:parameter_description}). Photons are fired perpendicularly, as an infinitely narrow beam onto the tissue, and are probabilistically propagated and absorbed. The back-reflected photons are then recorded and their ratio to the total number of fired photons gives spectral reflectance. However, MCML algorithm is not capable of simulating complex geometry or effects such as fluorescence. In this paper we consider only the tissue reflectance simulations. More complex simulation frameworks \cite{zhu2013review} can be used, to incorporate greater detail (and application specific characteristics) in the simulated model.

\subsubsection{Adapting to imaging system}
\label{sec:adapting_imaging_system}
Simulated spectral reflectances are an inherent property of the tissue, and independent of light source or acquisition system. Light exiting from the light source is reflected on the tissue surface, navigates through the optical components (including filters, lenses and optic fibers) and finally reaches the camera, to be transformed into electric current and measured as pixel intensities by the camera.


This section presents the transformation of spectral reflectances to values proportional to image intensities measured by a HSI camera. The camera intensity measurement $\mathbfcal{I}(\mathbf{t}_i) = [\mathcal{I}_1(\mathbf{t}_i), ...,  \mathcal{I}_\mathbf{B}(\mathbf{t}_i)]$\footnote[1]{we drop other variables in $\mathbfcal{I}(\cdot)$ to emphasize the pairwise relation with the tissue parameters $\mathbf{t}_i$} 
is dependent on factors presented in Tab. \ref{tab:camera_measurement_factors}. For large image intensities (such as those caused by shot noise) often Poisson distributed noise is used, but can be approximated by a multiplicative Gaussian\cite{healey1994radiometric}. Since in our current work, images are analyzed at a pixel level, geometric distortions due to lens effects are ignored. Combining the effects of all the components, intensity at a pixel is computed as,


\begin{equation}
\label{eq:image_intensity}
\mathcal{I}_j(\mathbf{t}_i) = w_j~\tau(\cdot) \int_{\lambda_{\text{min}}}^{\lambda_{\text{max}}} \xi_j(\lambda)\cdot \theta_j(\lambda)\cdot \mathbf{r}_{\text{sim}}(\lambda, \mathbf{t}_i)\, \textrm{d}\lambda  ~~~  \forall j \in [1,\mathbf{B}]
\end{equation}

where, $[\lambda_{\text{min}}, \lambda_{\text{max}}]$ is the range of bands for the chosen camera with $\mathbf{B}$ bands. Assuming that the camera response $\tau(\cdot)$ is linear and that the spatial inhomogeneties of the light source are independent of wavelength, $\xi_j(\lambda, \mathbf{p})=\alpha(\mathbf{p})\xi_j(\lambda)$ simplifies the model without loss in generality,

\begin{equation}
\label{eq:image_intensity_simplified}
\mathcal{I}_j(\mathbf{t}_i) =\alpha(\mathbf{p}) \cdot w_j \, \int_{\lambda_{\text{min}}}^{\lambda_{\text{max}}}\xi_j(\lambda)\cdot \theta_j(\lambda)\cdot \mathbf{r}_{\text{sim}}(\lambda,\mathbf{t}_i) \,\textrm{d}\lambda  ~~~  \forall j \in [1,\mathbf{B}]
\end{equation}

Tab. \ref{tab:camera_measurement_factors} explains the parameters in Eq. \ref{eq:image_intensity_simplified}. The factor $\alpha(\mathbf{p})$ accounts for constant multiplicative changes in reflectance (corresponding to the image brightness), caused by variation in distance or angle of the camera w.r.t the tissue and internal scaling of reflectance to pixels measured by a camera \cite{claridge2013model}.

The assessment framework thus outputs intensity values as measured by the virtual target camera for each tissue instance $\mathbf{t}_i$. The pairs, $[\mathbfcal{I}(\mathbf{t}_i), \mathbf{t}_i],~\text{where} ~~~ \mathbfcal{I}(\mathbf{t}_i) = [\mathcal{I}_j(\mathbf{t}_i)] ~~~ \forall j\in[1,\mathbf{B}]$, gives the tissue reflectance and the associated parameters, which can be used for metric computation described in the following section.

\subsection{Metric computation}
\label{sec:metric_computation}


The pair, $[\mathbfcal{I}(\mathbf{t}_i), \mathbf{t}_i]$, can be used by an engineer to determine the performance of the virtual camera (and the optical setup) in the context of a target application (to encode quality factors such as, sharpness, noise, dynamic range or contrast). It may, for instance be simply the SNR of the image or more complex measure such as, BRISQUE \cite{mittal2012no}. Alternately, to assess the performance in estimation of physiological parameters (Sec. \ref{sec:layered_tissue_model}), the user can choose a suitable inversion model (e.g. beer-lamberts \cite{zuzak2002visible} or a neural network \cite{zhang2010determination}) and use the simulated data (Sec. \ref{sec:generative_model}) to train and test the performance of the (virtual) optical system. To show how a metric can be chosen and used, we consider the example of estimating $s$.

The simulated image intensities $\mathcal{I}_j(\mathbf{t}_i)$ obtained from the previous section, depend on the spectral distribution of light reaching the sensor $(\xi_j(\lambda))$ and the image brightness $(\alpha(\mathbf{p}))$. To make the image data more robust to changes in $\xi_j(\lambda)$, each band intensity is divided by its volume $\mathbf{L}_j =\int_{\lambda_{\text{min}}}^{\lambda_{\text{max}}}\xi_j(\lambda)\, \textrm{d}\lambda$. 
\begin{equation}
\label{eq:band_reflectances}
r_{i,j}(\mathbf{t}_i) = \frac{\mathcal{I}_j(\mathbf{t}_i)}{\mathbf{L}_{j}}  ~~~  j \in [1,\mathbf{B}]
\end{equation}

The term $r_{i,j}(\mathbf{t}_i)$ is referred to as the camera band reflectance. Eq. \ref{eq:band_reflectances} corresponds to the calibration with a white target for real measurements \cite{lu2014medical}. The multiplicative factor $\alpha(\mathbf{p})$ in eq. \ref{eq:image_intensity_simplified} is accounted for by normalizing with the sum of all bands ($L_{1}$ normalization),

\begin{equation}
\label{eq:rn}
\tilde{r}_{i,j}(\mathbf{t}_i) = \frac{\alpha(\mathbf{p}) r_{i,j}(\mathbf{t}_i)} {\alpha(\mathbf{p})\sum_{k=1}^{\mathbf{B}} r_{i,k}(\mathbf{t}_i)} =\frac{r_{i,j}(\mathbf{t}_i)}{\sum_{k=1}^{\mathbf{B}} r_{i,k}(\mathbf{t}_i)}  ~~~  j \in [1,\mathbf{B}]
\end{equation}

Using the method proposed in Wirkert et al.\cite{wirkert2016robust}, we train a random forest regressor $(\hat{f})$ for inverting camera band reflectance $(\tilde{\mathbf{r}}_{i} = [\tilde{r}_{i,1}, \ldots \tilde{r}_{i,\mathbf{B}}])$ to estimate $s$. Here, $\mathbfcal{I}(\mathbf{t}_i)$ is the data and $\mathbf{t}_i$ (optically relevant tissue properties defined in Sec. \ref{sec:layered_tissue_model}) are the corresponding labels. The simulated data can be partitioned such that the training set is used for modeling $\hat{f}$ and the test set used to evaluate the model $\hat{f}$. For the example case, we define ``mean absolute error" in estimation as the performance metric;


\begin{equation}
\label{eq:absolute_error}
m(\Xi, \Theta, \mathcal{L}) = 1/N \sum_{i=1}^{N} |\hat{f}(\tilde{\mathbf{r}}_{i}(\mathbf{t}_i)) - s_{i}|
\end{equation}

The metric ($m$), is a function of all the optical components $(\Xi)$ in the setup, the light source $(\Theta)$ and the tissue $(\mathcal{L})$, which systematically quantifies the the performance to guide an engineer. Modifying each system component independently, such as bands acquired by the HSI camera, or adding a lens in the optical path, would modify the resulting camera reflectance $(\tilde{\mathbf{r}})$ and the estimated $s$ (for the example case); is quantified by $m$.





\section{Experiments and Results}
\label{sec:experiments_and_results}



This paper proposes a standardized framework the for the assessment of a HSI system in health-care applications. The approach relies on the computation of pairs of tissue parameters and resulting camera measurements $[\mathbfcal{I}(\mathbf{t}_i), \mathbf{t}_i]$ (as in Sec. \ref{sec:adapting_imaging_system}), which are fed into a user-defined metric for performance quantification as described in Sec. \ref{sec:metric_computation}. The data used for metric computation is obtained from the generative model (Sec. \ref{sec:generative_model}), which is the focus of our validation study and presented in Sec. \ref{sec:framework_verification}. Sec \ref{sec:sample_applications} demonstrates a use-case for evaluating commercial HSI cameras for estimation of $s$.

\subsection{Verification of generative model}
\label{sec:framework_verification}

The purpose of the first experiment was to assess whether the forward model produces realistic virtual measurements. For validation, we considered the three components of the generative model presented in Sec. \ref{sec:generative_model} separately.

\begin{figure}[h!]
        \centering
        \includegraphics[width=0.8 \columnwidth]{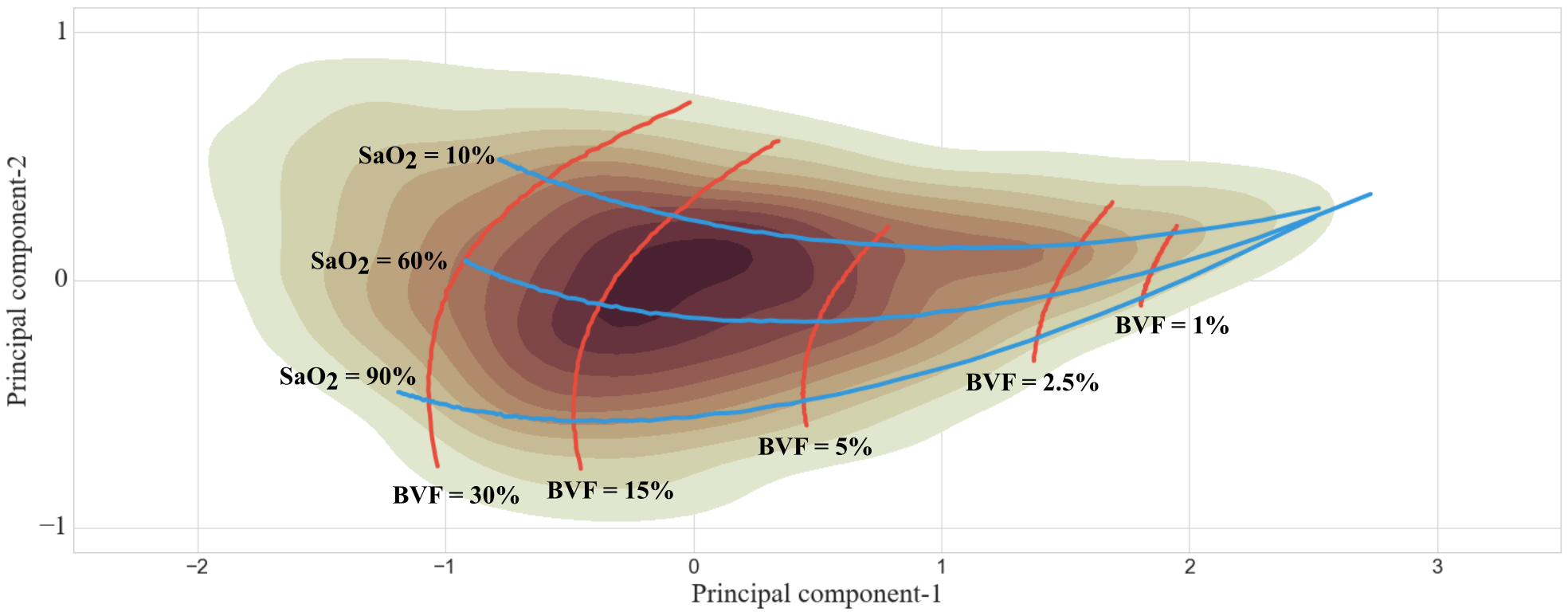}
        \caption{The grid indicates how varying $s$ and $v_{hb}$ changes the measurements in space spanned by the first two principal components of the simulated reflectances. Note that although the two principal components tend to follow the trend in varying $s$ and $v_{hb}$, these lines can not be directly interpreted as absolute values of $s$ and $v_{hb}$, because other factors such as scattering will cause movement on this simplified manifold.}\label{fig:general_data_plot_PCs}
    \end{figure}
\begin{figure}[h!]
        \centering
        \includegraphics[width=0.98 \columnwidth]{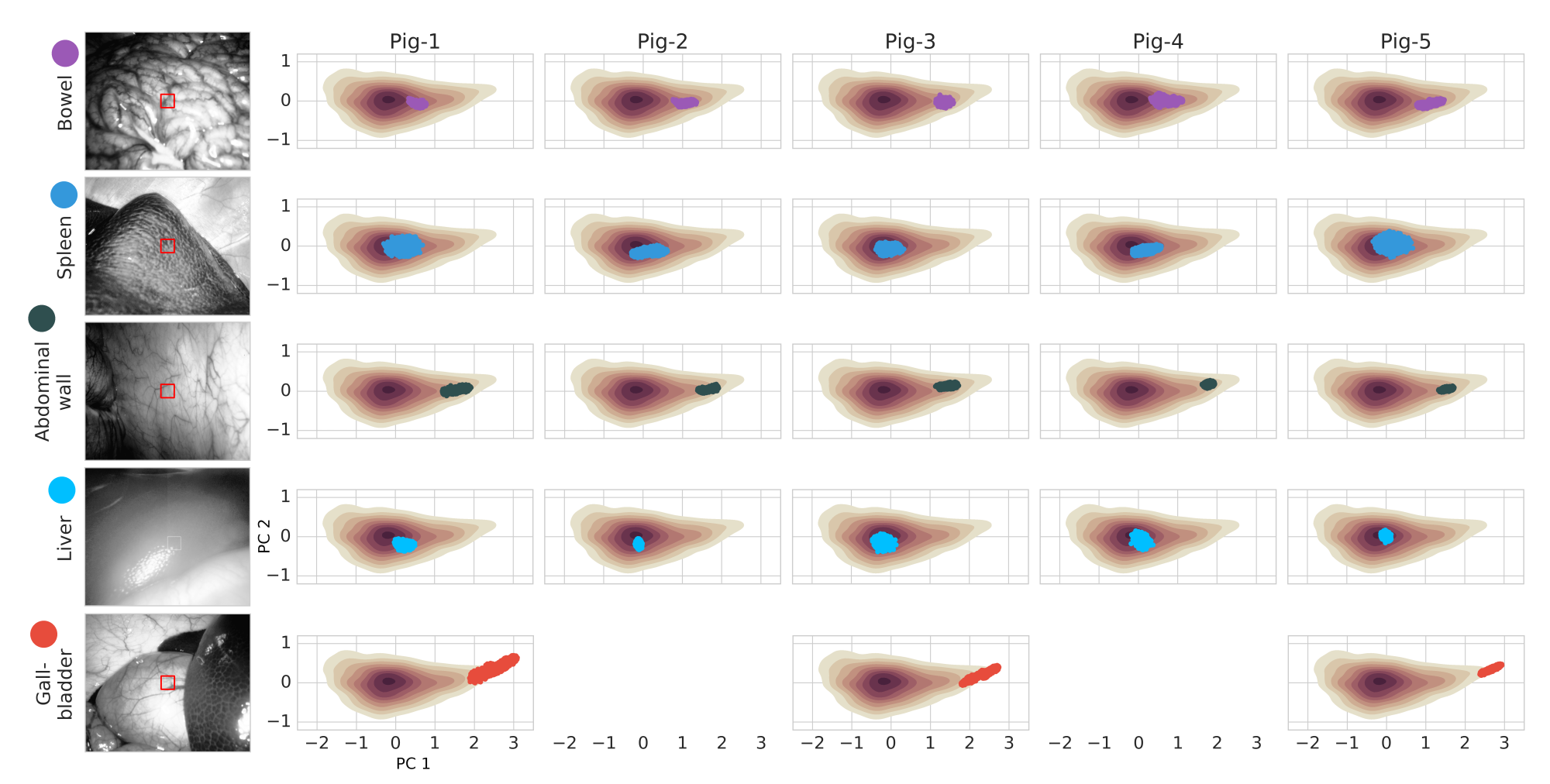}
        \caption{Five organs from five pigs projected onto the first two principal components of the simulated reflectance data plotted in brown. For pig-2 and pig-4, no gallbladder measurements were performed. The images on the left show the 560nm band recorded for the first pig. The depicted measurements are taken from the red rectangular region shown in the organ images. Except for gallbladder, all organs lie on the non-zero density estimates of the simulated data. Note that bile was not accounted for during simulations of the tissue model.}\label{fig:pig_matrix}
    \end{figure}
\begin{figure}[h!]
	\centering
	\includegraphics[width=0.95 \columnwidth]{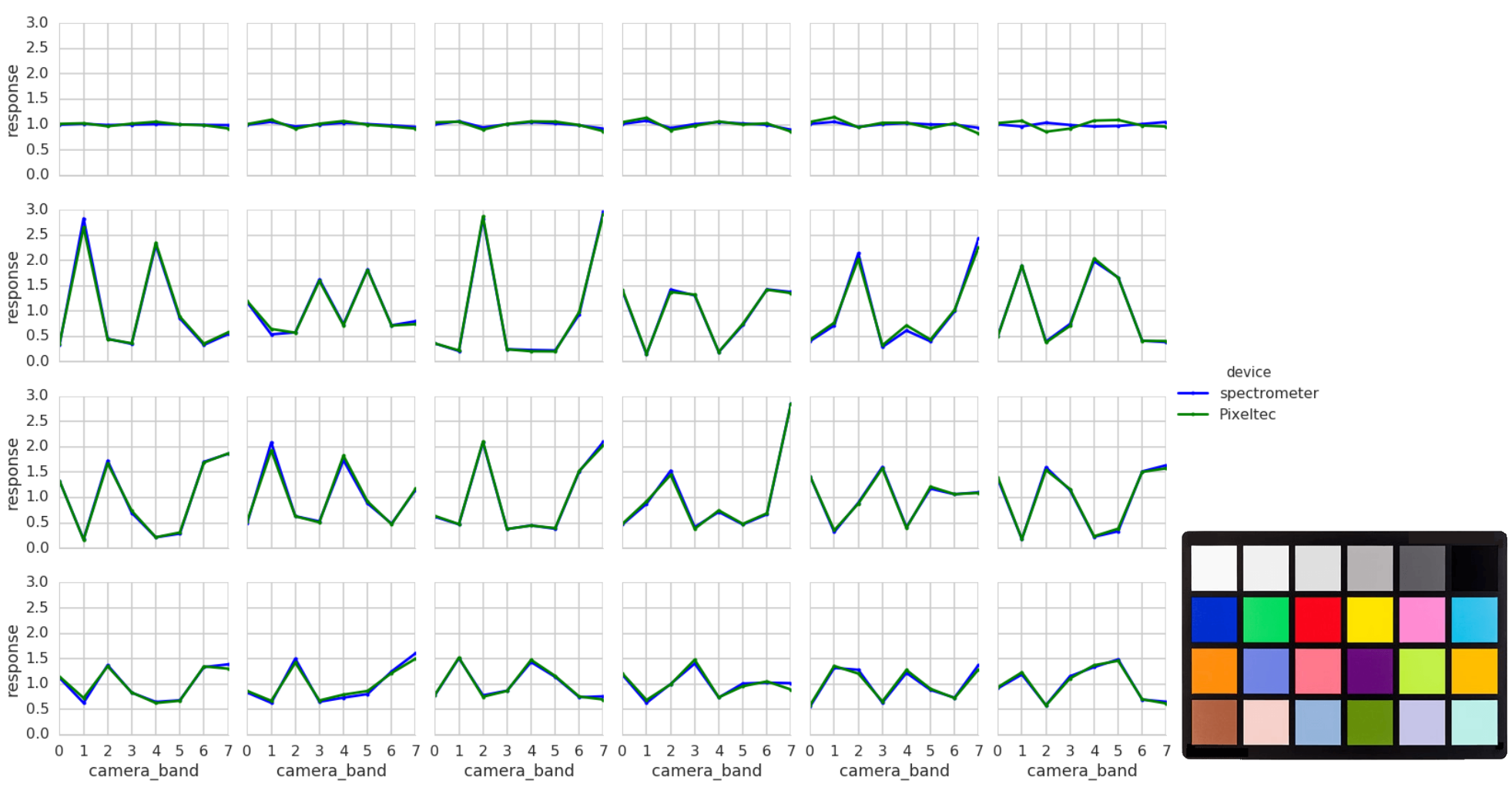}
	\caption{Color tiles measured with spectrometer and camera. The artificial optical system was applied to the spectrometer measurements to transform them into camera space. The figure on the bottom right shows the color tiles in the corresponding order of the plots.}\label{fig:color_tile_fit}
\end{figure}

\begin{itemize}
\item \textit{Camera model:} We start with the validation of the camera space transformation because, it will be used in the validation of the rest of the framework. Based on parameters defined in Sec. \ref{sec:adapting_imaging_system} the simulated reflectances are transformed to camera space. We propose to verify this formulation using an existing HSI setup that was presented in \cite{wirkert2016robust}. The optical setup consists of a Storz laparoscope, mounted on to Pixelteq SpectroCam\texttrademark~multispectral wheel camera, illuminated using a Storz Xenon Laparoscopic light source. The relative irradiance and the the transmissions of the laparoscope $(\xi_j(\lambda))$, the spectral responses of the filters of the camera were measured using a HR2000+ spectrometer (Ocean Optics, Largo, FL, USA) and verified with the data sheets provided by the manufacturer. All other parameters were obtained from the respective product data sheets. To validate if these measurements were correctly accounted in image system adaptation, we used color tiles\footnote[2]{X-Rite ColorChecker\textsuperscript{\tiny\textregistered} classic (Grand Rapids, MI, USA)} with uniform and accurate color representations. For each color tile, a spectrometer measurement was recorded and transformation as described in Sec. \ref{sec:adapting_imaging_system} was applied without adding noise. This was compared to the normalized mean of ten camera measurements. Fig. \ref{fig:color_tile_fit} shows the plots for all 24 color tiles. A mean error of -0.7\% and standard deviation of 5.4\% were observed. 

\item \textit{Light-tissue interaction:} Any simulation for light-tissue interaction can be employed to obtain realistic simulation of tissue reflectances in the proposed framework. In this paper we have chosen the GPU based MCML \cite{alerstam2010next}, which has been widely used and tested by the community and its verification will be omitted here. We refer the reader to previous work \cite{palmer2006monte,wilson2011models,hidovic2005modelling} for further study.

\item \textit{Tissue model:} Our pipeline for generating realistic camera measurements relies on a suitable (optical) model of tissue. To show that our assumptions on optical tissue properties are plausible, we extend the analysis presented in our previous work Wirkert et al. \cite{wirkert2017physiological}, which is based on a porcine study that involved taking real camera measurements from different organs and comparing them with the virtual measurements obtained from our framework. Fig. \ref{fig:general_data_plot_PCs} shows the contour plot of the first two principal components of the generalized reflectances (simulated using MCML with parameters in Tab. \ref{tab:da_generic}). Additionally, simulated data capturing variation of $s$ and $v_{hb}$ (while keeping the other parameters constant) is overlaid on the contour plot for intuitive understanding. In Fig. \ref{fig:pig_matrix} reflectances of five organs (bowel, spleen, abdominal wall, liver and gall-bladder) acquired from five pigs is projected on the generalized reflectance data. As detailed in Wirkert et al.\cite{wirkert2017physiological}, our study suggests that the simulated tissue reflectances can be used to analyze an HSI system for real-world cases, in case of unavailability of tissue specific real measurements.

\end{itemize}

\subsection{Sample Applications}
\label{sec:sample_applications}
We consider now the example case of $s$ estimation, to quantify performance of HSI cameras. The generalized tissue parameters shown in Tab. \ref{tab:da_generic} is used for simulating tissue reflectances. We use the random forest model introduced in Wirkert et al.\cite{wirkert2016robust} for inverting the simulated reflectances to estimate $s$. The ``mean absolute error" defined in Eq. \ref{eq:absolute_error} is used as the metric for performance quantification (as was described in Sec \ref{sec:metric_computation}). Using this metric, Sec. \ref{sec:camera_comparison} presents the framework-metric combination for comparison (and selection) of commercial (and other) HSI cameras. Sec. \ref{sec:camera_design} alternately presents an approach to design a custom virtual camera using the concepts of informative band selection tuned to $s$ estimation.

\subsubsection{Comparison of cameras for $SaO{_2}$ estimation}
\label{sec:camera_comparison}
We selected five commercial cameras (Table. \ref{tab:camera_list}) that can be used in laparoscopy; \begin{inparaenum}[(1)]
	\item Olympus Narrow band imaging (NBI) endoscope \cite{gono2004appearance},
	\item 3-band camera proposed by Kaneko et al.\cite{kaneko2014hypoxia} for measuring hypoxia in tissue,
	\item Pixelteq PixelCam\texttrademark~ with RGB + 3xNIR  \cite{pixelcam}.
	\item Pixelteq SpectroCam\texttrademark~multispectral wheel camera \cite{spectrocam}
	\item Ximea SNm4x4 hyperspectral snapshot mosaic camera \cite{ximea_SNm4x4}
\end{inparaenum}.

\begin{table}[!]
	\centering
	\caption{Cameras evaluated using the proposed framework and the filter responses are presented in Fig. \ref{fig:camera_filter_responses}.\label{tab:camera_list}}
	\begin{tabular}{llcl}
		\toprule
		Camera name                                                                                                                   & Short name   & \# of bands & Reference                                             \\ \midrule
		Olympus Narrow band imaging endoscope                                                                                         & NBI          & 2           & Gono et al.\cite{gono2004appearance} \\ \midrule
		3-band camera for hypoxia                                                                                                     & HypoxiaCam  & 3           & Kaneko et al.\cite{kaneko2014hypoxia} \\ \midrule
		Pixelteq PixelCam\texttrademark~with RGB + 3xNIR            & Pixelcam     & 6           & PixelCam\cite{pixelcam}              \\ \midrule
		Pixelteq SpectroCam\texttrademark~multispectral wheel camera & Spectrocam   & 8           & SpectroCam\cite{spectrocam}          \\ \midrule
		Ximea SNm4x4 hyperspectral snapshot mosaic camera                                                                             & Ximea SNm4x4 & 16          & Ximea\cite{ximea_SNm4x4} \\ \midrule
		Virtual camera with 16-equispaced normal bands & VirtualCam & 16 & \\ \midrule
	\end{tabular}
\end{table}

\begin{figure}[!tbp]
	\centering
	\includegraphics[width=\textwidth]{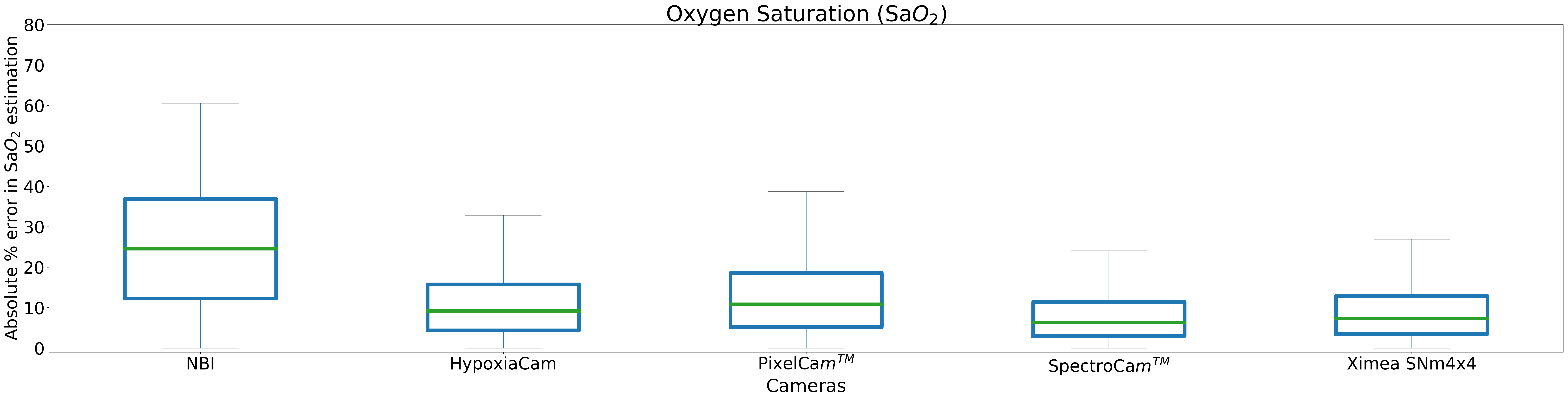}\\
	\caption{Plots showing the absolute error in estimation of $s$ for various commercial and research cameras.}\label{fig:sao2_evaluation}
\end{figure}

We observe from Fig. \ref{fig:sao2_evaluation} that, for general tissue representation, the $[$median error and std dev.$]$ for estimation of $s$ using Pixelteq SpectroCam\texttrademark~ \cite{spectrocam}, HypoxiaCam \cite{kaneko2014hypoxia} and NBI camera \cite{gono2004appearance} and  were $[$6.3\%, 9.2\%$]$, $[$9.1\%, 10\%$]$ and $[$24.6\%, 14.5\%$]$ respectively. We observe that the 3-band HypoxiaCam provides a performance comparable to the 8-band SpectroCam\texttrademark~ \cite{spectrocam}, and the 2-band NBI camera \cite{gono2004appearance} does not encode sufficient information about $s$. The 16-band Ximea SN4x4 with the imec snapshot sensor, which is comparatively lower (spatial) resolution sensor to others in the list, results in a performance $[$7.3\%, 9.2\%$]$ comparable to SpectroCam\texttrademark~ in a smaller form-factor for estimation of $s$.

\begin{figure}[!tbp]
	\centering 
	\includegraphics[width=\textwidth]{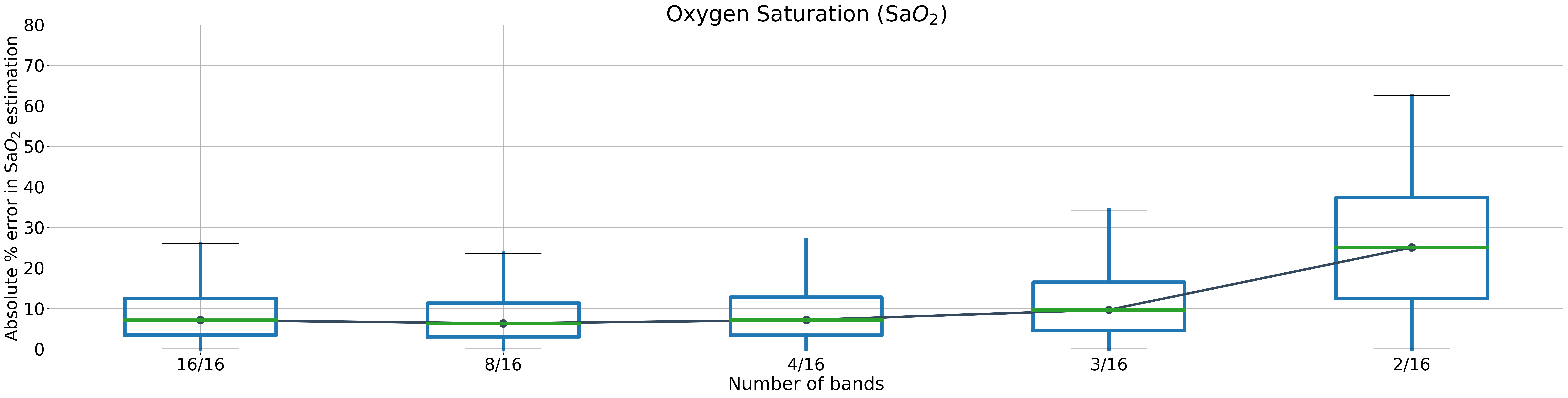}\\
	\caption{Figure shows the plot for absolute error in estimation of $s$, while successively eliminating the least informative bands quantified using feature importance values obtained from random forest regressor in sklearn\cite{scikit-learn}. Here results from $[16, 8, 4, 3, 2]$ bands have been shown.}\label{fig:camera_design}
\end{figure}

\subsubsection{Camera Design}
\label{sec:camera_design}

Informative band selection is cruicial step during the design of a custom camera suited to a target application, to improve the discriminative power of the measured signal. For a HSI camera, we can select the most informative bands that would for example, increase the predictive capability of tissue physiological parameters ($s~\text{or}~v_{hb}$). Several algorithms for band selection have been proposed in literature \cite{wood2008optimal,wirkert2014endoscopic}. Our framework is compatible with such techniques, hence the user can perform a complete analysis within this environment whilst designing a suitable camera and choosing the optical system components.

In this section we perform a na\"{i}ve band selection from a 16-band virtual camera, by selecting those bands which have the most informative quantified by the ``\textit{feature importances}'' obtained from the random forest regressor \cite{breiman2001random} (using the `scikit-learn' module\cite{scikit-learn}). Fig. \ref{fig:camera_design} shows the ``mean absolute error'' in estimation of $s$, while successively removing the least informative bands at each step. We observe that the mean and the standard deviation of the absolute error varies from $[7\%, 9\%]$ (for 16 bands) to $[7\%, 9.6\%]$ (for 4 bands) to $[25\%, 14\%]$ (for 2 bands). In the context of translatable clinical application a 4-band camera delivers a better cost to performance ratio for the engineer. The analysis exemplifies the simplicity of quantifying the application-specific camera design process using the proposed framework.

\section{Discussion}
\label{sec:discussion}

In this paper we have presented a quantification framework for assisting the research community and system engineers alike in choosing, designing or optimizing HSI cameras and the optical system, for health-care applications. The primary advantage of the proposed tool is that, no physical device setup is necessary, since the assessment is performed in a virtual environment. The parameters needed as user input the proposed framework, can typically be obtained from product data-sheets or from the manufacturer upon request. 

In analyzing the proposed framework, we have firstly, in Sec. \ref{sec:framework_verification}, validated the generative model as the core of our approach by showing that, \begin{inparaenum}[(1)]
\item real tissue measurements resemble simulated measurements produced with our framework and that
\item the conversion of reflectances to camera measurements is valid.
\end{inparaenum} Then in Sec. \ref{sec:camera_comparison}, we demonstrated how the commercial cameras can be compared within the proposed framework, while selecting oxygenation estimation as an example target application and suitably choosing a performance metric. Using the same application-metric combination, Sec. \ref{sec:camera_design}, showed how the work-flow of the proposed assessment framework complements informative band selection. Here using a 16-band virtual camera, feature importance (for each band) estimated by the random forest regressor was used for determining the most informative bands.

A generalized multi-layered tissue model presented in Table. \ref{tab:da_generic} was considered in the example cases in Sec. . A well-performing camera on a generic dataset may be interesting, but it is bound to contain simulated reflectances not relevant to the target clinical application (such as for skin or brain tissue in case of abdominal surgery). To apply the analysis on a target organ, the user may consider using the transfer learning approach presented in Wirkert et al.\cite{wirkert2017physiological}, in which, (unlabeled) images from the target organ (obtained using the desired imaging setup) can be used to weight the generalized data set to obtain organ-specific samples that can be used for analysis.

We have limited our analysis to using hemoglobin as the major chromophore. Other chromophores (such as bile or melanin) can be included in the simulation within the tissue configuration described in Sec. \ref{sec:layered_tissue_model} and the sample file shown in Fig. \ref{fig:tissue_ini_sample}. Simulating complex geometry or effects such as fluorescence, algorithms such as Geant4 \cite{allison2016recent} may to be used. Owing to the modular formulation of our framework, this can be easily incorporated. As another possible use case, the proposed framework can be used to determine the best light source for a given application by selecting the metric $m$ (Eq. \ref{eq:absolute_error}) as a function of $\Xi, \Theta, \mathcal{L}$ defined in Tab. \ref{tab:camera_measurement_factors} and \ref{tab:parameter_description}.

In health-care applications, there is strong price-performance trade-off for clinical translation of a medical device. An important consideration in the design of a HSI camera is tradeoff between spectral and spatial resolution. For example, Sec. \ref{sec:camera_comparison} showed that the fewer bands selected by Kaneko et al.\cite{kaneko2014hypoxia}, give a performance close to a high spectral resolution camera \cite{ximea_SNm4x4}, albeit with a more complex design. Fewer bands lead to faster acquisition times, a smaller form-factor for the clinical device and (possibly) higher spatial resolution, all of which are important factors in clinical applications.

A weakness of our framework would be the hypothesis that the simulations reflect realistic scenarios. We have presented an evaluation to perform a qualitative comparison on real and simulated reflectances in Sec. \ref{sec:framework_verification}. Fig. \ref{fig:pig_matrix} shows the projection of 5 organs from 5 different pigs on our generalized reflectance data. MC simulations speeded-up by the GPU still takes a long time, for example, it took us $(\sim$ 10 days on a single Titan X GPU to generate  $\sim 500k$ samples. However, it should be noted that the most time-consuming part of the generative model is the generation of tissue reflectances from the layered tissue model. As the conversion of reflectances to virtual camera measurements is fast, testing of various camera configurations can be done efficiently once a sufficiently large dataset of tissue reflectances has been generated. 

It should further be noted that the presented generic framework is applicable for image level analysis too. However, MCML (the tool used in this paper) currently simulates only individual pixels with the assumption that adjacent pixels are independent and uncorrelated. The next step for improving this framework would be to provide an approach to generate small patches of realistic tissue images that can be considered for 2D analysis.

In conclusion, we hope that this generalized framework can assist researchers and industry experts alike in simplifying the process of designing (or choosing) a hyperspectral camera and the associated optical system in most health-care application.

\begin{figure}
    \centering
    \subfloat[Olympus NBI gastroscope\cite{gono2004appearance}]{\includegraphics[width=0.45\textwidth]{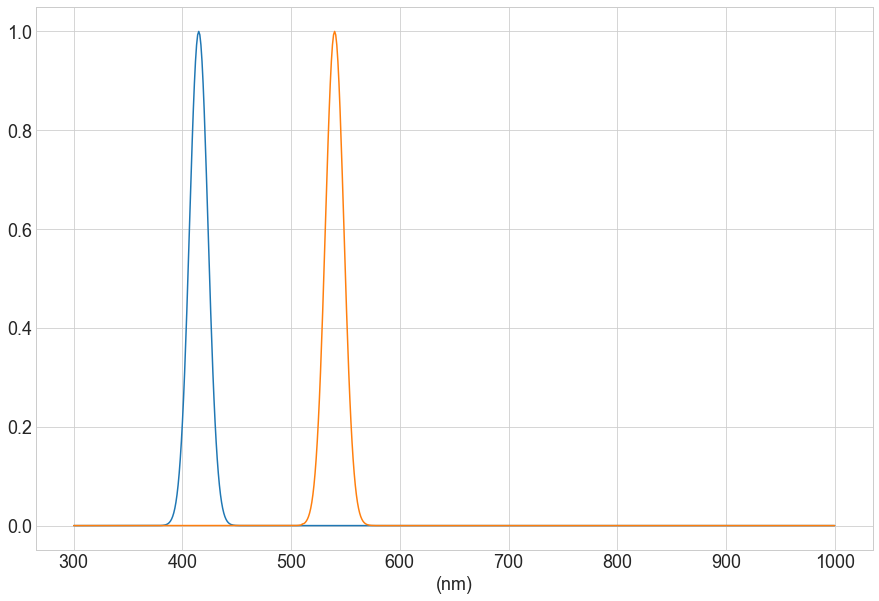}\label{fig:commercial_NBI}}
    \subfloat[The filter responses obtained from Kaneko et al.\cite{kaneko2014hypoxia} - HypoxiaCam ]{\includegraphics[width=0.45\textwidth]{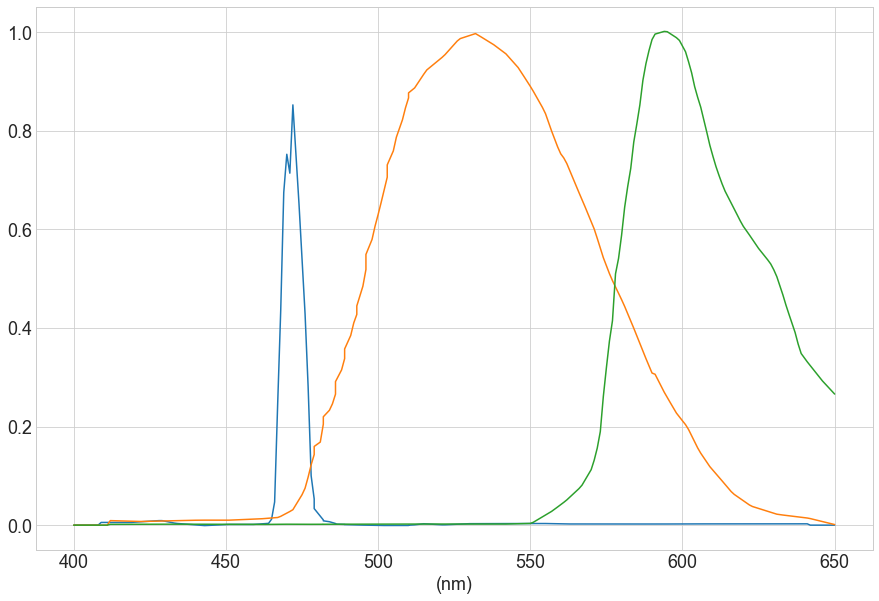}\label{fig:hypoxiacam}}\\
    
    \subfloat[PixelCam\texttrademark~ - RGB+3xNIR \cite{pixelcam}]{\includegraphics[width=0.45\textwidth]{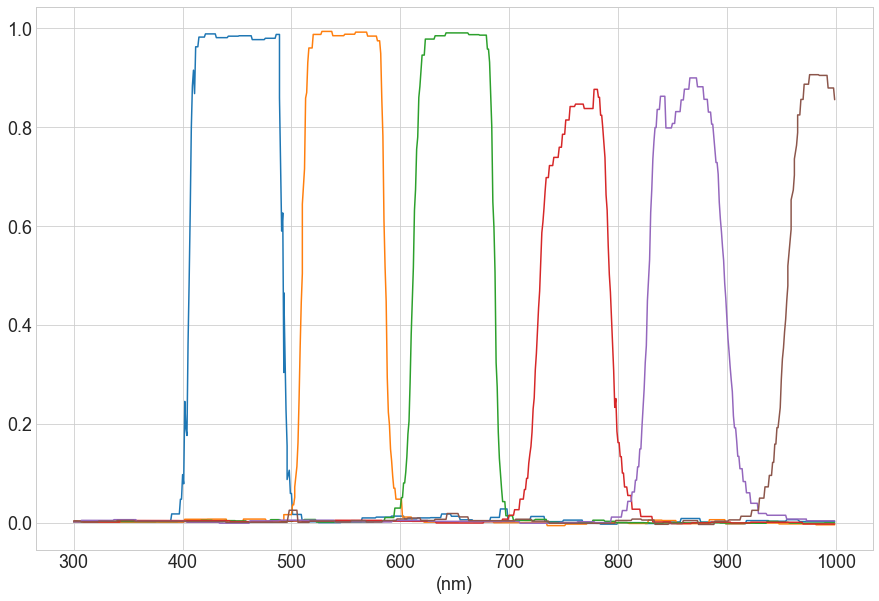}\label{fig:commercial_PixelCam_RGB3xNIR}}
    \subfloat[SpectroCam\texttrademark~-8 bands \cite{spectrocam}]{\includegraphics[width=0.45\textwidth]{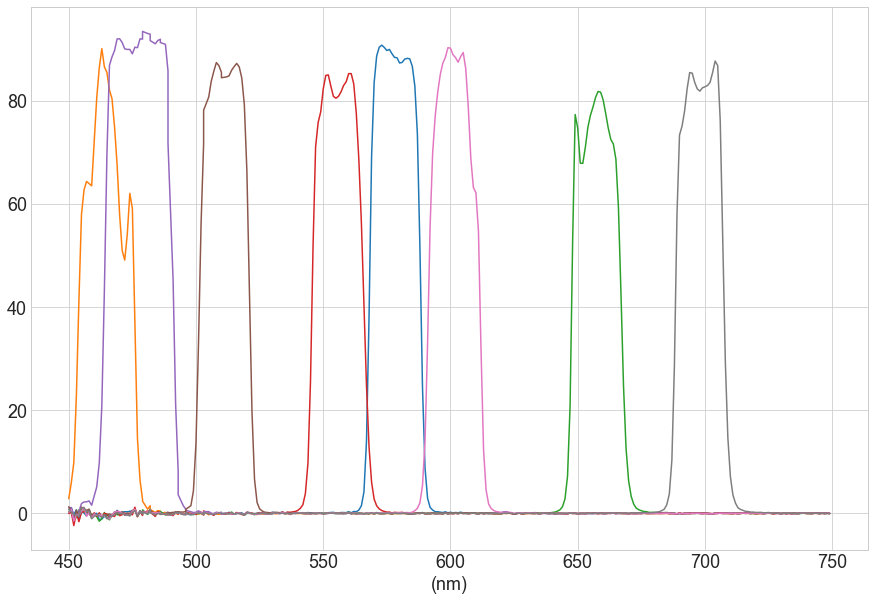}\label{fig:commercial_spectrocam_8}}\\
    
    \subfloat[Ximea SNm4x4 xiSpec with imec imaging chip \cite{ximea_SNm4x4}]{\includegraphics[width=0.45\textwidth]{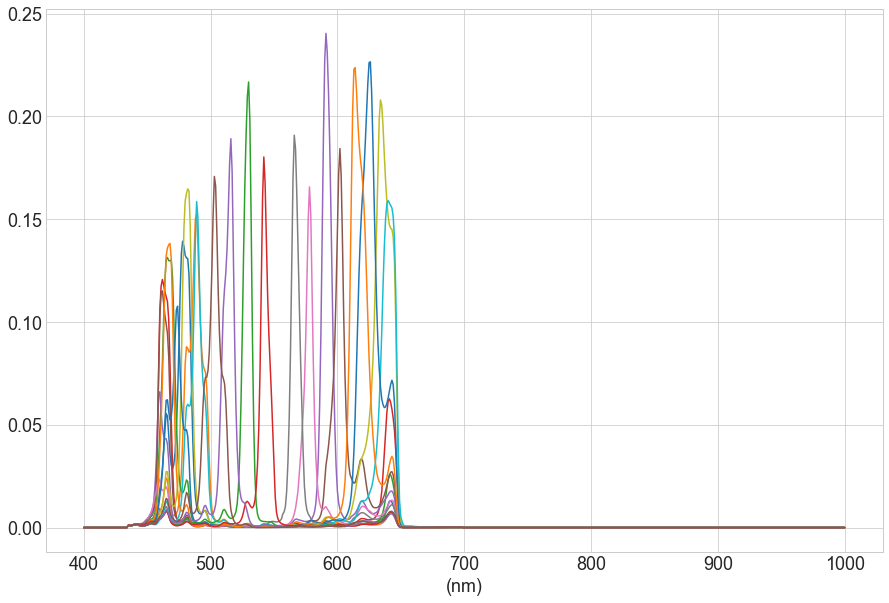}\label{fig:commercial_ximea}}
    \subfloat[VirtualCam with 16 equispaced normal distributed bands]{\includegraphics[width=0.45\textwidth]{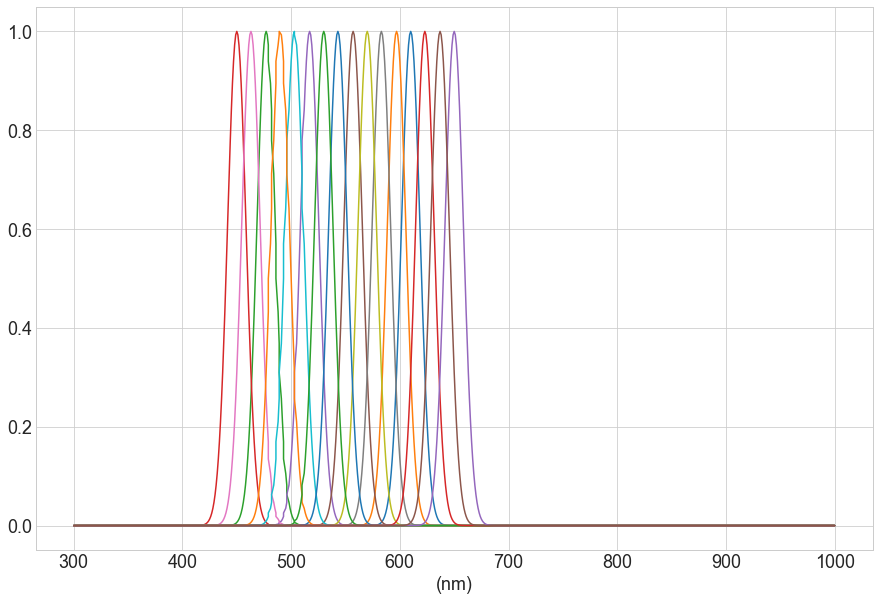}\label{fig:commercial_PixelCam_RGBNIR}}\\
    
    \caption{Filter responses of cameras selected for analysis in this paper.}\label{fig:camera_filter_responses}
\end{figure}

\acknowledgments     
 
The authors would like to acknowledge support from the European Union through the ERC starting grant COMBIOSCOPY under the New Horizon Framework Programme under grant agreement ERC-2015-StG-37960.

\section*{Disclosures}

There are no financial conflicts of interest to disclose.


\bibliography{main}   
\bibliographystyle{spiejour}   

\end{spacing}
\end{document}